# Modern Location-based Service Technologies: Visible Light Positioning

Shangsheng Wen, Yingcong Chen

*School of Materials Science and Engineering, South China University of Technology, Guangzhou, China (e-mail: shshwen@scut.edu.cn)*

*Abstract*— **With the development of wireless communications and the increasing computing power of variety mobile devices, LBS (Location Based Service) technologies getting more and more attention as it can provide most flexibility and convenience in modern people's life. For this survey, we will first give a comprehensive introduction about LBS, including definition, advantages, application, and potential privacy problem. Then, we will present more detailed discussion focusing on the location technologies which is an essential part in LBS framework.**

*Keywords*— **Location Based Service, Location technology, Indoor Position Systems, Visible Light Positioning**

## I. INTRODUCTION

LBS (Location Based Service) refers to those service which can functioning easily based on its described location with the aid of different kinds of indexing and guidance services. [1] Normally, it will first determine the location of the mobile devices or users, then provide variable kinds of information service related to location. Therefore, LBS is to use the Internet or wireless network, between fixed users or mobile users, complete the two functions of positioning and service. [2] This kind of service, not only make it easier for those users to achieve more pertinent and latest information about their neighbour situations, but also make the businesses to supply recent updates to clients more practicable. [3]

### A. Main Types

There are mainly 2 types of LBS technology, which is Pull LBS, push LBS and Tracking LBS.

Pull (Reactive or user requested) LBS conveys data specifically asked for, from the client. Clients start correspondence by initiating a service request to the LBS provider. In view of the location information provided, the service provider answers with service content.

Push (Proactive or Triggered) LBS initiated by an event, which could be activated if a particular region is entered or by a periodic timer. More specifically the service provider activates the transfer of the asked service. It puts forward value-added services in return of user's location information.

Tracking LBS refers to those service providers who persistently tracks the clients. Under this kind of LBS, Ceaseless location information of entity is recorded where entity could be human or non-human (e.g., commodities) and the like.

### B. Advantages

Compared with traditional service technology, benefits of LBS for both users and providers are shown as discussed as following.

*1) Benefits for Users:* First, it can help the users to pick useful information thus make quick decision based on the location information given by the mobile devices with the users, such as help the visitors to picking the best restaurant in a specific range or finding out available time slot for a new movie in adjacent silver screen. Second, it can provide easily neglected but of vital importance tips for the users in time and precisely. For example, when there is an unpredicted traffic jam, instant warning or suggestion can be given based on LBS. Third, it is efficient for providing specific, as LBS can acquire the location data and other relevant information from sensors on smart adaptable gadgets. In other words, clients do not need to manually input information every time they want to get access to certain kinds of service.

*2) Benefits for Service Providers*: First, more service can be provided with the location information provided by senior users. Second, Application can update automatically more easily. As LBS help with the collecting location data along with tagged data, it will be more convenient for a application inside a mobile device to get updated with those reference bringing an umbrella given the weather forecast for a specific region can be presented.

### C. Application and Marketing

Internet and robust location technologies such as GPS, LBS provides users with a variety of location-related services rather effectively. The above characteristics help LBS attract more and more attention in recent years [4] . There are several areas show big potential for applicating LBS technologies [5].

According to study, location-based service's application areas can be broadly classified in following 4 kind: Emergency Service, Informational Services News, Tracking Services and Entertainment Services. And detailed function in different application areas is shown in Fig.1.

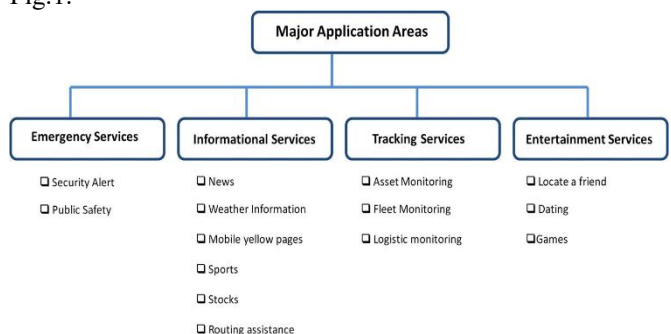

Fig. 1 Major LBS application areas

As for marketing performance, based on the survey conducted in 2020, the location-based services market size was valued at $28.95 billion in 2019, and is projected to reach $183.81 billion by 2027, growing at a CAGR of 26.3% from 2020 to 2027.

Besides, the key players operating in the location-based services market include Alcatel-Lucent SA, Apple, Inc., AT&T Inc., Bharti Airtel, LTD., Cisco Systems, Inc.,

Google Inc., HERE, International Business Machines Corporation, Microsoft Corporation, Oracle Corporation, and Qualcomm Inc. These players have adopted various strategies to expand its business, strengthen its product portfolio, and increase their LBS market penetration, which, in turn, is expected to support the global location-based services market growth.

### D. Existing Problems and Solutions

Although LBS can bring a lot of convenience also help to both clients and service providers, the data security becomes a new concern when it comes to LBS.

*1) Privacy Concern:* According to the working principle, users need to provide their location to service providers to obtain services, so the location information in the process of sending may be intercepted by the attacker, resulting in the user's location or identity information disclosure. For example, If the user initiates the query continuously in a period, and the attacker continuous monitoring, it may be inferred that the user's trajectory or the location of the next moment and other important information, to the user's personal and property security poses a serious threat [5].

*2) Latest solutions:* Many privacy preserving schemes have been proposed to protect location privacy. Dargahi et al. employ k-anonymity method to ensure that an LBS provider cannot distinguish target user from at least k other users and thus prevent user's location from compromising. Ying et al. propose dynamic Mix-Zone for location privacy in vehicular networks. Zhu et al. propose an efficient query scheme in which LBS provider's data is outsourced to cloud server in an encrypted manner. LBS users encrypt their queries using the improved homomorphic encryption technique. All these solutions can provide state of art performance [6].

## II. FRAMEWORK AND LOCATION TECHNOLOGIES USED IN LBS

### A. LBS Architecture

Typical LBS architecture requires five basic components as shown in Fig. 2.

1. The portable mobile devices play important roles for the user to submit the required service information. The results of the requested service can be given by speech, text, pictures, and so on.

2. Communication network works as a spine for the whole LBS framework, which exchanges the client information and service request from the mobile devices to the location-based service provider and subsequently the requested services from service provider to the client.

3. The location-based service provider offers different services to the client after service request query processing. Such services can be used as different functions, for example: estimating the localization in a worldwide setting as navigation, path planning for finding a suitable route, finding the "bad" information to a specific spot or discovering specific data of the particular objects which is the interest of clients.

4. Data and content provider is used to keep user's location information and the related privacy materials.

5. The positioning technologies is one of the fundamental components in LBS, which is adopted to track the mobile client's visits and to convey required information assistance at the related time and related location to the correct user. Therefore, the compelling and the dominant use of positioning technologies significantly affect the performance of the LBS. In LBS, the positioning technology is used to register mobile location movement. Talking about the positioning technology, the most remarkable ones is the Global Positioning System (GPS). However, GPS is only available in outdoor scenery. For indoor application, there are also many potential techniques such as Bluetooth, WLAN, RFID, UWB, Wi-Fi, IR, Zigbee, and so on [7] . In the next section, we would like to introduce some modern positioning technologies, especially the indoor positioning system (IPS).

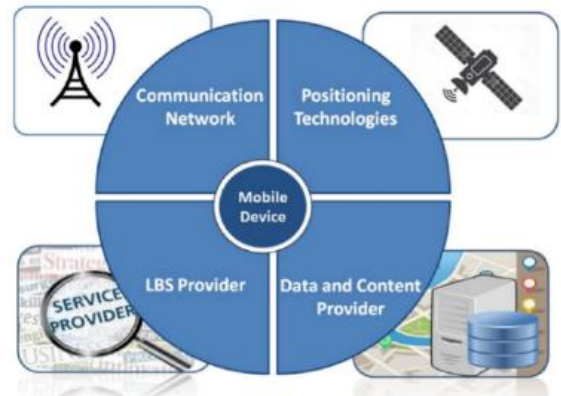

Fig. 2 The architecture of the LBS

### B. Modern Positioning Technologies

Positioning is a crucial component in LBS as it allows get information about the location of the mobile device or the clients. The development of satellite systems (such as GPS, GNSS) revolutionized the accuracy of location information, however for many LBS applications the GPS-positioning accuracy level is more than enough. What's more, indoor scenery is an especially challenging problem, since the localization cannot be achieved by GPS due to the satellite signal being greatly attenuated [8]. Fortunately, the current popularization of the indoor mobile positioning applications seems to bring perspectives of a significant value especially for department stores, shopping malls, airports etc. In this section, we would like to introduce the modern positioning technologies.

*1) GPS:* GPS has been extensively used for positioning in outdoor environments such as in car navigation, mobile phones, ships, planes, surveying of public works and so on. Any device with a GPS receiver (includes most smartphones) can ping the satellites with that receiver. This will cause it to communicate with at least four satellites, and the satellites can compare the signal delay to pinpoint where the signal originated. GPS-based LBS is the simple and standard solution. Sony Ericsson's "Near Me" is one such example. It is used to maintain knowledge of the exact location, however can be expensive for the end-user, as they would have to invest in a GPS-equipped handset. GPS is based on the concept of trilateration, a basic geometric principle that allows finding one location if one knows its distance from other, already known locations.

*2) WLAN:* Wireless local area networks (WLANs) have been successfully deployed in our daily live, which can provide internet wireless access services by installing

WLAN access points (APs) in public hotspot areas, such as airports, coffee shops, and conference centers. When a mobile device detects the Wi-Fi signal, preinstalled positioning software scans it and compare to reference database, then based on the strength of one or several signals, it calculates user's location.

*3) Bluetooth:* Bluetooth offers low-cost, energy-efficient and easy to deploy indoor positioning and tracking technology. This helps to easily locate or track items and people, find directions and other important information within buildings and facilities such as airports, shopping malls, and others. Millions of people rely on Bluetooth item finding solutions to track down lost keys and other personal items. Bluetooth indoor positioning systems (IPS) are helping millions of travelers navigate through busy airports and shoppers through malls. And warehouses and facilities around the globe are using Bluetooth real time locating systems (RTLS) to better manage resources and optimize the flow of materials.

*4) RFID:* RFID tracking is effectively a combination of these other methods. The RFID scanner typically has a static location. By pinging off of other networks, the location of the scanner can be logged. When the RFID scanner is activated, it can tag its location when it records the access. This can be used to identify the location of the device accessing the scanner.

*5) Ultra-wideband (UWB):* Ultra-wideband is a short-range radio technology which can be used for indoor positioning. In contrast to Bluetooth Low Energy and Wi-Fi, positioning is done with transit time methodology (Time of Flight, ToF) instead of the measurement of signal strengths (Receive Signal Strength Indicator, RSSI). This method measures the running time of light between an object and several receivers. For the exact localization of an object at least 3 receivers are necessary (trilateration). Also, there must be direct line-of-sight between receiver and transmitter.

However, these positioning systems have the following disadvantages: (1) these systems need a lot of infrastructure investment, which increases the cost of operating and the complexity of control; (2) since the wireless signal has uneven spatial distribution and low stability, there is a strong volatility at the same positioning location point, which results in the poor accuracy of indoor positioning; (3) they also add electromagnetic (EM) interference, which decreases the positioning accuracy severely when the system is shared by multiusers, in addition, they are not applicable in the no EM interference scenes, such as hospitals, airports, and so on.

*C. New Generation IPS: VLC*

Due to the high transmission capacities, optical wireless networks play an important role in our modern information technology. Advances in visible light communication (VLC) technology and the ubiquity of illumination facility have led to a growing interest in VLC-based LBS (also name as visible light positioning, VLP). The VLC using Light Emitting Diodes (LEDs) has become a hot topic in wireless communications. The White-LED lighting technology as a type of green lighting technology is considered as the next generation indoor lighting because of the advantage of high luminous efficiency (up to 110lm/W), long service life (up to 50000hours) and fast response speed (up to 10MHZ). Distinct from the older illumination technologies, the LED switches to intensity levels at a very fast rate which is imperceptible by human eyes. Hence, data can be encoded in the emitting light in various ways which mean LEDs can provide communication/positioning and illumination, simultaneously.

Compared with the traditional indoor positioning technology, VLP has the advantages of high positioning accuracy, no EM interference, fewer additional modules, good security, and the possibility of combining communication with lighting, thus it has aroused the attention of many experts and scholars in the world. The comparison between VLP and traditional IPS can be seen as Table I. Comparing the trade-off between the overall performance and the complexity of the IPSs, we believe that the VLP should be most potential one.

TABLE I THE COMPARISON AMONG DIFFERENT POSITIONING TECHNOLOGIES

| System | Range | Accuracy | Scalability | Cost | Power Consumption | Advantages | Constraints |
|---|---|---|---|---|---|---|---|
| GPS | 16km | 6-20m | Low | High | High | Convers entire earth | Expensive infrastructure; Only outdoor |
| FM | 100km | 2-4m | Low | Low | Low | Less susceptible to objects | Signal changes little in small distance |
| Cellular Networks | 80km | 2.5-20m | Low | Medium | High | Widely covered scope | Low accuracy |
| ZigBee | 30-60m | 1-10m | Low | Low | Low | Low cost and power consumption | Low accuracy; low data transmission |
| Wi-Fi | 35m | 1-5m | Medium | Medium | High | Low cost | If fingerprint is used, environment dependent and high deployment cost for building database |
| Infrared | Few meters | 1-2m | Low | Medium | Low | Low and power consumption | Short range; cost for extra hardware |
| Ultrasound | Tens of meters | 0.01-1m | Low | Medium | Low | Good accuracy | Sensitive to environment; cost for extra hardware |
| Bluetooth | 10 m | 1-5m | High | Low | Low | Low cost and power consumption | Cover range is limited; limitation in user mobility |
| UWB | Few meters | 0.01-1m | Low | High | Low | Good accuracy | Cover range is limited; high cost; problems in non-line of light |
| RFID | 1m | 1-2m | Medium | Low | Low | Low cost and power consumption | Response time is high; limitation in user mobility |
| VLP | 3-5m | 0.03-0.5m | Low | Low | Very low | High accuracy positioning and illumination, simultaneously | Need line-of-sight; need modification on LED |

The modulated LED broadcasts its unique identity (ID) via switching at a high frequency imperceptible to the human eye, but which can be recognized by photodiodes (PD) and rolling shutter effect (RSE) cameras. The LED ID can be mapped/modulated once for all since they are normally fixed and not easily vulnerable to environmental changes. Hence, the "Last Mile Problem" of localization is solved via VLP and the pre-built LED feature map. The VLP methods can be broadly divided into three categories: scene analysis, proximity, and triangulation.

Scene analysis refers to a group of positioning algorithms that matches measured information to a pre-calibration database to realize positioning, thus omitting the computation process. However, this kind of positioning algorithm requires accurate pre-calibration for a specific environment and cannot be instantly deployed in a new setting.

The proximity method is a very simple positioning algorithm that it relies only upon a grid of reference points, each having a known position; when a positioning terminal collects a signal from a single source, it is considered to be co-located with the source. Therefore, the accuracy is no more than the resolution of the grid itself.

Triangulation is a kind of localization algorithm that uses the geometric properties of triangles to estimate the position, and it is mainly divided into two branches: the angulation and the Lateration [9]. Angulation is used to measure the arrival angle of the receiver, which is relative to a certain position reference point. Location estimation is then performed by finding intersection of direction lines. Angulation method is essentially an imaging positioning technology and captures the field of view of the camera. Some state-of-the-art (STOA) camera-based VLP systems can offer centimeter-level accuracy on commodity smartphones [10] or mobile robots [11]. While the Lateration method estimates the specific coordinates of a receiver by measuring the horizontal distances between the receiver and the multiple positioning reference points. The location distance is measured indirectly by measuring the received signal strength (RSS), time of arrival (TOA), or time difference of arrival (TDOA). Apart from providing high accuracy indoor localization, VLC-based LBS can also provide high-speed data access for the user [12]. Each LED access point can provide not only unique ID address, but also Li-Fi network to transmit information. We believe that, with the development of the technology, the VLC-based LBS might become one of the popular technologies in LBS and the internet of things (IoT).

III. OUR WORK IN VLP

We start our research in the field of VLC and VLP since 2014, we have published many research articles in prestigious international journals and conferences. The representative works from our group can be seen in Ref. [13-17]. A blog[1] demonstrates the application of VLP in robotics and IoT from our group, and 24 authorized patents can be seen in Ref. [18-41]. A presentation about our work in the field of high accuracy robot localization using VLP can also be seen in[2]. For the readers that get insights from our research works, please cite our works.

IV. CONCLUSIONS

The papers explored the LBS infrastructure, advantages, application, and potential privacy problem. What is more, we analysis the modern indoor positioning technology and introduce the next generation indoor positioning technology —— visible light positioning. We believe that, in the future, the light-based IPS can provide broad application prospects.

---

[1] http://t.csdn.cn/Os3YW
[2] https://b23.tv/FWQWPsg